\documentclass[10pt,conference]{IEEEtran}
\IEEEoverridecommandlockouts
\let\oldLetter\Letter  
\usepackage{ifsym}
\let\Letter\oldLetter  

\usepackage{cite}
\usepackage{amsmath,amssymb,amsfonts}
\usepackage{graphicx}
\usepackage{textcomp}

\usepackage{xspace}
\usepackage{booktabs}
\usepackage{multirow}
\usepackage[table,xcdraw]{xcolor}
\usepackage{caption}
\usepackage[utf8]{inputenc}
\usepackage[switch]{lineno}

\usepackage{wrapfig}
\usepackage{tabularx}
\usepackage{makecell}
\usepackage{multirow}
\usepackage{subfigure}
\usepackage{algorithm}
\usepackage{algorithmicx}
\usepackage{amsmath,amsfonts}
\usepackage{amssymb}
\usepackage{array}
\usepackage{balance}
 \usepackage{booktabs}
\usepackage[skip=1pt,labelfont=bf]{caption}
 \usepackage{calligra}
\usepackage{color}
\usepackage{colortbl}
\usepackage{courier}
\usepackage{csvsimple}
\usepackage{enumitem}
\usepackage{fancybox}
\usepackage{fontenc}
\usepackage{graphicx}
\usepackage{listings}
\usepackage{longtable}
\usepackage{lscape}
\usepackage{makecell}
\usepackage{marvosym}
\usepackage{moreverb}
\usepackage{multicol}
\usepackage{multirow}
\usepackage{pifont}
\usepackage{rotating}
\usepackage{setspace}
\usepackage{subfigure}
\usepackage[most]{tcolorbox}
\usepackage{threeparttable}
\usepackage{tikz}
\usepackage{soul}
\usepackage[normalem]{ulem}
\usepackage{url}
\usepackage{soul}
\usepackage{wasysym}
\usepackage{xspace}
\usepackage{amsthm}
\usepackage{hyperref}

\usepackage{todonotes} 
\usepackage{colortbl}
\usepackage{pifont}
\usepackage{xspace}
\usepackage{tcolorbox}
\usepackage{enumitem}

\usepackage{stfloats}

\algnewcommand\algorithmicforeach{\textbf{for each}}
\algdef{S}[FOR]{ForEach}[1]{\algorithmicforeach\ #1\ \algorithmicdo}

\newcolumntype{L}[1]{>{\raggedright\let\newline\\\arraybackslash\hspace{0pt}}m{#1}}
\newcolumntype{C}[1]{>{\centering\let\newline\\\arraybackslash\hspace{0pt}}m{#1}}
\newcolumntype{R}[1]{>{\raggedleft\let\newline\\\arraybackslash\hspace{0pt}}m{#1}}

\definecolor{codegreen}{rgb}{0,0.6,0}
\definecolor{codered}{rgb}{1,0,0}
\definecolor{codegray}{rgb}{0.5,0.5,0.5}
\definecolor{codepurple}{rgb}{0.58,0,0.82}
\definecolor{backcolour}{rgb}{0.95,0.95,0.92}
\definecolor{lightgray}{gray}{0.9}

\lstdefinestyle{mystyle}{
    commentstyle=\color{codegreen},
    keywordstyle=\color{magenta},
    numberstyle=\small\color{black},
    stringstyle=\color{codepurple},
    basicstyle=\scriptsize\ttfamily,
    breakatwhitespace=false,
    breaklines=true,
    captionpos=b,
    keepspaces=true,
    showspaces=false,
    showstringspaces=false,
    showtabs=false,
    tabsize=2
}

\lstdefinelanguage{diff}{
  morecomment=[f][\color{blue}]{@@},     
  morecomment=[f][\color{red}]-,         
  morecomment=[f][\color{codegreen}]+,       
  morecomment=[f][\color{red}]{---}, 
  morecomment=[f][\color{codegreen}]{+++},
}

\setlist{noitemsep} 

\definecolor{darkpastelred}{rgb}{0.76, 0.23, 0.13}
\definecolor{ao(english)}{rgb}{0.0, 0.5, 0.0}

\definecolor{darkpastelred}{rgb}{0.76, 0.23, 0.13}
\definecolor{ao(english)}{rgb}{0.0, 0.5, 0.0}

\definecolor{yellow}{RGB}{255,255,153}
\definecolor{grey}{RGB}{224,224,224}

\newboolean{showcomments}
\setboolean{showcomments}{true}
\ifthenelse{\boolean{showcomments}}
 { \newcommand{\mynote}[2]{
      \fbox{\bfseries\sffamily\scriptsize#1}
        {\small$\blacktriangleright$\textsf{\emph{#2}}$\blacktriangleleft$}}}
        { \newcommand{\mynote}[2]{}}

\definecolor{DarkOrange}{rgb}{0.8,0.3,0.0}
\definecolor{DarkCyan}{rgb}{0.0, 0.55, 0.55}
\definecolor{DarkCyel}{rgb}{1.0, 0.49, 0.0}
\definecolor{yellow-green}{rgb}{0.6, 0.8, 0.2}

\newcolumntype{?}{!{\vrule width 1pt}}

\definecolor{todogreen}{rgb}{0.0, 0.5, 0.0}

\usepackage{times}
\usepackage{latexsym}
\newcommand{\find}[1]{
\begin{tcolorbox}[leftrule=0.2mm,toprule=0mm,bottomrule=0mm,left=0.0pt,right=0pt,top=0pt,bottom=0pt]
\em #1
\end{tcolorbox}
}

\newcommand{\tool}{\textit{\texttt{ExGroFi}}\xspace}

\def\BibTeX{{\rm B\kern-.05em{\sc i\kern-.025em b}\kern-.08em
    T\kern-.1667em\lower.7ex\hbox{E}\kern-.125emX}}
\begin{document}


\title{Delving into Commit-Issue Correlation\\to Enhance Commit Message Generation Models}

\makeatletter
\newcommand{\linebreakand}{%
  \end{@IEEEauthorhalign}
  \hfill\mbox{}\par
  \mbox{}\hfill\begin{@IEEEauthorhalign}
}
\makeatother

\author{
    \IEEEauthorblockN{1\textsuperscript{st} Liran Wang}
    \IEEEauthorblockA{\textit{Beihang University} \\
    Beijing, China \\
    wanglr@buaa.edu.cn}
\and
    \IEEEauthorblockN{2\textsuperscript{nd} Xunzhu Tang}
    \IEEEauthorblockA{\textit{University of Luxembourg} \\
    Luxembourg City, Luxembourg \\
    xunzhu.tang@uni.lu}
\and
    \IEEEauthorblockN{3\textsuperscript{rd} Yichen He}
    \IEEEauthorblockA{\textit{Beihang University} \\
    Beijing, China \\
    hyc2026@buaa.edu.cn}
\and
    \IEEEauthorblockN{4\textsuperscript{th} Changyu Ren}
    \IEEEauthorblockA{\textit{Beihang University} \\
    Beijing, China \\
    cyren@buaa.edu.cn}
\linebreakand
    \IEEEauthorblockN{5\textsuperscript{th} Shuhua Shi}
    \IEEEauthorblockA{\textit{Beihang University} \\
    Beijing, China \\
    shishuhua@buaa.edu.cn}
\and
    \IEEEauthorblockN{6\textsuperscript{th} Chaoran Yan}
    \IEEEauthorblockA{\textit{Beihang University} \\
    Beijing, China \\
    ycr2345@buaa.edu.cn}
\and
    \IEEEauthorblockN{7\textsuperscript{th} Zhoujun Li*}
    \IEEEauthorblockA{\textit{Beihang University} \\
    Beijing, China \\
    lizj@buaa.edu.cn}

\thanks{*Corresponding author}
}

\maketitle
\thispagestyle{plain}
\pagestyle{plain}

\begin{abstract}
Commit message generation (CMG) is a challenging task in automated software engineering that aims to generate natural language descriptions of code changes for commits. Previous methods all start from the modified code snippets, outputting commit messages through template-based, retrieval-based, or learning-based models. While these methods can summarize what is modified from the perspective of code, they struggle to provide reasons for the commit. The correlation between commits and issues that could be a critical factor for generating rational commit messages is still unexplored.

In this work, we delve into the correlation between commits and issues from the perspective of dataset and methodology. We construct the first dataset anchored on combining correlated commits and issues. The dataset consists of an unlabeled commit-issue parallel part and a labeled part in which each example is provided with human-annotated rational information in the issue. Furthermore, we propose \tool (\underline{Ex}traction, \underline{Gro}unding, \underline{Fi}ne-tuning), a novel paradigm that can introduce the correlation between commits and issues into the training phase of models. To evaluate whether it is effective, we perform comprehensive experiments with various state-of-the-art CMG models. The results show that compared with the original models, the performance of \tool-enhanced models is significantly improved.

\end{abstract}


\IEEEpeerreviewmaketitle



\section{Introduction}




Commit messages are in the format of natural language for summarizing and explaining the intention of commits during the maintenance of software projects~\cite{buse10, cortes14}. Thus, a well-written commit message can help code reviewers quickly understand what the commit is and why the commit is proposed without any detailed look into the complex codes~\cite{mockus00, what_why}.

However, commits are often complex, making it difficult and error-prone to summarize with a concise commit message in manual ways~\cite{what_why},~\cite{boa},~\cite{nmt},~\cite{maalej10}. According to Loyola et al.~\cite{nmt}, 16\% messages in a widely used dataset CommitGen~\cite{commitgen} are noisy, which means they contain little information about the commit. In addition, in the situation of increasingly fast-paced modern software development, manually writing high-quality commit messages will also be time-consuming~\cite{mcmd}. 

To simplify the process of writing commit messages, researchers have proposed various techniques for generating commit messages automatically. Since the original input and target output of commit message generation (CMG) are in different modalities, code and natural language, most CMG approaches formulate the CMG pipeline as a translation task.

Mainstream generation methods can be categorized into template-based~\cite{buse2010automatically, cortes14, shen2016automatic}, retrieval-based~\cite{nngen, cc2vec}, learning-based~\cite{commitgen},\cite{nmt},\cite{codisum},\cite{ptr},\cite{coregen},~\cite{fira} and hybrid models~\cite{atom},~\cite{race},~\cite{corec}. Template-based techniques start from parsing the modification of source code and then generate commit messages with pre-defined rules. Retrieval-based models calculate a similarity score between the requested input and other code changes in the training set. Then, they utilize a ranking algorithm to select the code change with the highest similarity and use its corresponding commit message as output. Learning-based models first use a trained encoder to embed the input code change into semantic space. Then, a decoder or pointer network will be used to generate output in natural language space according to the representation of code change provided by the encoder. Hybrid models combine the characteristics of learning-based and retrieval-based methods, such as RACE~\cite{race}, which proposed a retrieval-augmented mechanism that can be leveraged in the generation phase. For learning-based and hybrid approaches, the process of embedding may also need some additional parsed structure information such as abstract syntax tree~\cite{atom}. Additionally, with the development of pre-trained language models in natural language processing~\cite{li2021pretrained, qiu2020pre}, more and more pre-trained code change representation methods have emerged~\cite{codebert},~\cite{codetrans},~\cite{codet5}. This also significantly improves the quality of embedding and makes excellent progress in CMG tasks.

Although existing approaches show promising performance, they still tend to generate meaningless and irrational commit messages\cite{corec}. An apparent restriction is that all these techniques only receive input information directly or indirectly from the code itself. While the code, as the object of modification, can usually only reflect the result of the commit rather than the reason. This makes it difficult for existing methods to generate good commit messages that contain both summary (i.e., What) and motivation (i.e., Why) of the change~\cite{what_why}. In fact, a lot of commit messages are highly related to the content of some issues or bug reports, which usually describes the detailed reason for the commit~\cite{anvik06, zhu16}. As shown in Figure~\ref{motivation}: By reading the given issue, a developer can understand that there is a bug to be fixed because of ``the incorrect result of MATCH with eval''. The title of the issue summarizes the problem in a few words, and the body of the issue elaborates on the problem in detail by describing actual behaviors and expected behaviors. In the commit message, the keywords to the issue (``MATCH'' and ``eval'') reemerge. Upon resolution of the issue, an event established a correlation between the commit and the issue. 
None of the existing research on CMG tasks has focused on this correlation between commit and issue. Therefore, exploring this correlation in-depth and leveraging it to improve the performance of CMG approaches has become the primary motivation of this paper.

\begin{figure}[!t] 
	\centering
	\includegraphics[width=1\linewidth]{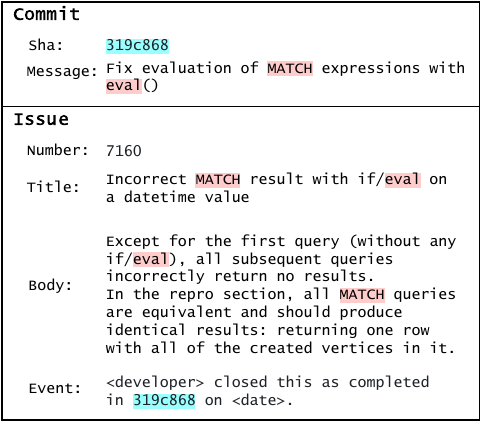} 
	\caption{Correlation between commit and issue.}
	\label{motivation}
\end{figure}

In this work, we explored the correlation between commits and issues from the perspective of data and approach. First, to fill the gap in the research community on this topic, we collect a large commit-issue parallel dataset to understand what kind of rational information issues can provide. While constructing the dataset, we also consider identifying fine-grained knowledge in the data. Therefore, the whole dataset consists of two parts. One part is unlabeled, and the other is labeled in which each example is provided with human-annotated rational information in the issue. Further, from the perspective of approach, we propose a novel training paradigm \tool (\textbf{E}xtration, \textbf{G}rounding, \textbf{F}ine-tuning) that can improve the performance of pre-trained CMG models by introducing the correlation between commit and issue into the training phase. The paradigm consists of three pipelined stages. The first stage aims to extract fine-grained knowledge from issues. Then the second stage leverages the extracted knowledge to better represent the code change into semantic space. In the last stage, a fine-tuning task is performed to generate commit messages.

To evaluate whether \tool is effective for CMG task, we perform an extensive evaluation with the collected dataset. The overall experiment result shows that the performance of state-of-the-art CMG models can be significantly improved by introducing the correlation between commit and issue using \tool. Further, the result of two fine-grained experiments can prove the effectiveness of each stage of \tool . In addition, considering the incompleteness of automatic metrics, we also perform a human evaluation from four aspects: rationality, comprehensiveness,  conciseness, and expressiveness. Results show that \tool can improve the rationality of the generated commit messages while maintaining the level of other indicators.

In summary,  the main contributions of this paper are:
\begin{itemize}
    \item A high-quality commit-issue parallel dataset, which is the first dataset that can be used to explore the correlation between commit and issue.
    \item A novel paradigm \tool for commit message generation, which consists of an extraction stage, a grounding stage, and a fine-tuning stage. The paradigm introduces the extracted rational information from issues to the training phase of commit message generation models.
    \item A comprehensive evaluation of \tool, including a comparison study, a verification of the performance of each stage, and a human evaluation. All results suggest the effectiveness of our paradigm from different perspectives.
\end{itemize}

The remainder of this paper is organized as follows. Sec.~\ref{Sec.2} elaborates on the background of this work. Sec.~\ref{Sec.3} introduces the collected commit-issue parallel dataset. Sec.~\ref{Sec.4} explains in detail the proposed \tool paradigm. Sec.~\ref{Sec.5} and Sec.~\ref{Sec.6} present our experimental configuration and analyze the results. Sec.~\ref{Sec.7} discusses the threat of validity and limitations of our work. Sec.~\ref{Sec.8} lists some related works. Sec.~\ref{Sec.9} gives a conclusion to this paper.
\section{BACKGROUND}
\label{Sec.2}
Issues are a great way to keep track of what needs to be done and what has been done in a project~\cite{bettenburg2008makes}. It can always guide developers in writing concise and descriptive commit messages. Therefore, commits and issues are always highly correlated~\cite{quatrain},~\cite{vieira19},~\cite{zhou2017automated}. However, none of the existing automatic commit message generation approaches leveraged this correlation. In this section, we further analyze the reasons for this situation and lead to the motivation for our work.

\subsection{Correlation between Commits and Issues}
Both issues and commits play a crucial role in collaborative software development. An issue refers to a bug report, a feature suggestion, or any problem that needs to be addressed for a project. It can be identified by any developer and user of the software, and is typically tracked in a management tool like Jira~\cite{jira}, Trello\cite{trello}, or GitHub~\cite{github}. A commit is a specific change made to the source code. It includes a brief message that describes the changes made, along with any relevant metadata such as the author, timestamp, and the specific files and lines of code that were modified. Issues often drive the creation of commits~\cite{what_why}. When developers start working on an issue, they will make one or more commits to the codebase to resolve the issue. 

In the scenario of writing a commit message, the developer will briefly describe the content of the modification according to the code change and attach an explanation by referencing an issue. One common way to relate a commit message to an issue is to reference the issue number or link in the message explicitly~\cite{linux_kernel, sawant18}. For example, the commit message might include a line like  ``Refactor code related to issue 1234" or ``Fixes issue \underline{\#1234}". In other cases, the developer may also directly extract some sentences from the issue as the reason~\cite{pits, secret}. Therefore, there should be a strong correlation between the text of issues and commit messages. Tian et al.~\cite{quatrain} has validated the existence of this correlation from the perspective of semantic similarity. They used BERT~\cite{bert} to embed text in the semantic space. Moreover, the results show that the original issue (bug report) and associated commit message pairs are much more similar than the randomly selected ones. Consequently, it is easy to imagine that issues will contain much rational information for writing commit messages.

\subsection{Lack of Data and Research}
\label{Sec.2.2}

Although the correlation between commits and issues is obvious, there is still no research that exploits this kind of knowledge to solve the CMG task. This situation is mainly due to the difficulty in obtaining high-quality issue data. As the open-source software (OSS) community continues to thrive~\cite{emerge, zhu16} , an increasing number of OSS projects are turning to public issue-tracking systems (ITS), such as Jira~\cite{jira} and the issue-tracking functionality of Github~\cite{github_issue} to manage issues. This enables timely feedback on problems in the project but also results in a diverse and inconsistent quality of issue data due to the following barriers. (1) Technical knowledge. Developers with different expertise levels may have different abilities to describe complex issues~\cite{developer, li2022identifying}. This can lead to some developers writing more complex and accurate issues, while others may have difficulty communicating the full extent of the problem or suggestion being reported. (2) Motivation. Developers with different motivations may have different levels of interest in contributing to the project and writing high-quality issues~\cite{breu2009frequently}. This can lead to some developers carefully research and report well-written issues, while others may not be as invested in the project and may write lower-quality issues. (3) Communication skills. Developers with different communication skills may have different abilities to clearly and concisely describe an issue~\cite{commenters}. This can lead to some developers writing more organized and easy-to-understand issues, while others may have difficulty expressing themselves in a way that is clear and concise.

Therefore, the current situation is that although a large amount of open-source issue data exists, there is no high-quality and sufficiently large dataset that provides commits with related issues, nor is there a sophisticated paradigm that can effectively leverage the correlation between commit and issue. This leads to the motivation for our work: construct a high-quality commit-issue parallel dataset and explore methods to efficiently use commit-issue correlation for commit message generation.
\section{THE DATASET}
\label{Sec.3}
To investigate the significance of the commit-issue correlation for the CMG task, we construct a commit-issue parallel dataset. This dataset fills a gap in the academic community in this area and may provide a reasonable basis for future research. It provides both raw unlabeled corpus and labeled data that can guide models to learn from external knowledge. In this section, we first introduce the construction process of the dataset. Then we explain how we define the information that should be labeled and the annotation details. Finally, some useful statistics and an example are presented.

\subsection{Construction}

\subsubsection{Collection}
Considering the ease of access and volume of data, we collect data from Github~\cite{github}, a web-based platform for version control and collaboration. GitHub not only has a large number of high-quality open-source projects, but also has a comprehensive issue-tracking functionality~\cite{github_issue}. We only focus on projects developed in Java because it is one of the most widely used programming languages in the industry and is the most studied in academic research~\cite{mcmd}. We retrieve data from the top 1,000 Java repositories according to the number of stars using PyGithub~\cite{pygithub}, the Python package of Github REST API~\cite{githubrestapi}. 

Our purpose is to construct a dataset that provides linked commit-issue pairs in real software projects. However, the original GitHub REST API does not provide an interface that can directly fetch this kind of data. Le et al.~\cite{rclinker} proposed a discriminative model to predict if a link exists between a commit message and a bug report. However, this model was trained on generated commit messages and bug reports, which means it can not ensure the objectivity and truthfulness of data. After an in-depth examination of the GitHub REST API, we find that it provides a class of IssueEvent object that can help us reconstruct the needed information. A Github IssueEvent refers to an activity related to an issue. It occurs when an issue is opened, closed, edited, or referenced. For each project issue, we filter out all the commits that reference it by specifying the type of IssueEvent as ``referenced". At the same time, we maintain a mapping from commit to a list of issues, and every time we get a commit that references an issue, we append that issue to the corresponding list in the mapping. In this way, we successfully collect all the primary data we needed, and we keep the following information for further processing:
\begin{itemize}
    \item \textbf{Commit message}: The whole commit message of the commit.
    \item \textbf{Issues}: A list of issues that are referenced by the commit. For each issue, its title (a summary of the issue in a few words) and its body (a detailed description of the issue) are provided.
    \item \textbf{Files}: A list of files that are modified in the commit. For each file, we keep its relative path in the project (contains its filename) and code change (formatted as the output of the \textit{git diff} command).
\end{itemize}

\subsubsection{Text processing}
The text data obtained from the GitHub REST API usually contains a lot of noise~\cite{nngen, what_why, mcmd}. This makes it challenging to extract valuable insights from them without proper text-processing techniques. In our dataset, there are three types of text data: commit message, issue title, and issue body. Commit messages and issue titles are relatively easy to process because they are usually short. While for issue bodies, the situation is more complex. As free-form content, issue bodies can include plain text, markdown, and images. It should provide all the necessary information to understand and reproduce the issue, including steps to reproduce, expected behavior, actual behavior, error messages, example code snippets, and other relevant information~\cite{bettenburg2008makes}. In addition, as mentioned in Sec.~\ref{Sec.2.2}, different proposers have different writing habits. Therefore, the quality of issue bodies varies. In order to obtain a more generalized knowledge from issues, we need to normalize their content. To do this, we respectively replace URLs and code snippets in issue bodies with special tokens ``[URL]" and ``[CODE]". Moreover, for issue numbers mentioned in commit messages, we replaced them with the token ``[ISSUE\_NUMBER]".

\subsubsection{Filtering}
Not every example that we obtain can be deposited into the dataset. We also perform further filtering on the processed text data. According to Liu et al.~\cite{nngen}, there will be some commit messages that are automatically generated by bots or trivial that contain little and redundant information. Following their work, we also remove the data containing these commit messages and only keep English data. In addition, considering the length limitation of the existing pre-trained models on the input token sequence, we also filtered the data based on the length of tokenized sequences. Data containing fields that exceed the length limit (1024 tokens) are also excluded.

\subsection{Annotation for Issues}
\subsubsection{Definition}
\label{Sec.3.2.1}
As mentioned in Sec.~\ref{Sec.2.2}, the writing style and text length of issues proposed by different people vary greatly. This also makes it challenging to explore the correlation between commits and issues. Even if we carefully process the text and eliminate the content that does not contain semantic information as much as possible, it is still difficult to directly use it to provide valuable information for models. Therefore, we try to manually annotate the fine-grained information in issues such that it can help models to learn more knowledge and better embed code change in semantic space. Observing a large amount of data, we define two types of fine-grained information as follows.
\paragraph{Issue type}
Issues can be categorized into different types based on motivation~\cite{behavior}. GitHub issues themselves do not have predefined types or topics. However, project maintainers can use a flexible labeling system to categorize issues based on their nature, priority, or other relevant criteria. By applying labels, maintainers can create a more organized issue-tracking system that is easier to navigate and manage. According to labels, we summarize three commonly used issue types:
\begin{itemize}
    \item \textbf{Bug report}: A bug report is an issue that describes unexpected behavior or an error in the code. It needs to be fixed to ensure the proper functioning of the software.
    \item \textbf{Feature request}: A feature request is a suggestion or proposal for a new feature or functionality that users would like to see in the software. The way users use the software will change because of the commits related to it.
    \item \textbf{Enhancement}: An enhancement is a suggestion to improve an existing feature or functionality of the software. Unlike feature requests, it usually does not result in a change in the way users use the software. Common enhancement includes performance optimization, compatibility optimization, etc.
\end{itemize}

\paragraph{State information}
\label{state_info}
Issues always contain some descriptive content of the current situation and expected results~\cite{bettenburg2008makes}, and these descriptions are the most critical content to tell developers why the modifications should be done. We define this kind of knowledge as actual/expected state information. By analyzing the actual state and comparing it with the expected state, developers can make better changes to the code. Since the objective of a commit is to update the current codebase to meet expectations, its commit message should be highly related to the state information we defined. From the perspective of representation, the actual/expected state can also be regarded as the projection of the code before/after the commit in semantic space. This means that the text containing actual/expected states extracted from issues could be used as a supervised objective for code change representation learning.

\subsubsection{Annotation}
According to our definition of fine-grained information in issues, we perform an annotation of the collected data. We use Label Studio~\cite{label_studio} as the annotation platform. The commit message of a commit, together with the title and body of all issues related to this commit, are shown on the labeling page of each example. Five well-trained annotators were asked to label the type of each issue and to extract actual state expected state information from it. We finally obtain 960 labeled examples.

\subsection{Stratification}
The whole dataset has two parts: 
\begin{itemize}
    \item The base part consists of all unlabeled commit-issue data. This part can be regarded as a CMG dataset, but for each commit, all related issues are attached.
    \item The fine part consists of all labelled commit-issue data. Each of the data contains not only the related issue list but also the type and the actual/expected state information of each issue.
\end{itemize}
For the base part, we randomly stratify it into training, validation and test split with a ratio of 8:1:1. The base part contains 19262 data entries in total, and the fine part comprises 960 data entries.

\begin{figure}[!t]
    \centering
    \includegraphics[width=1\linewidth]{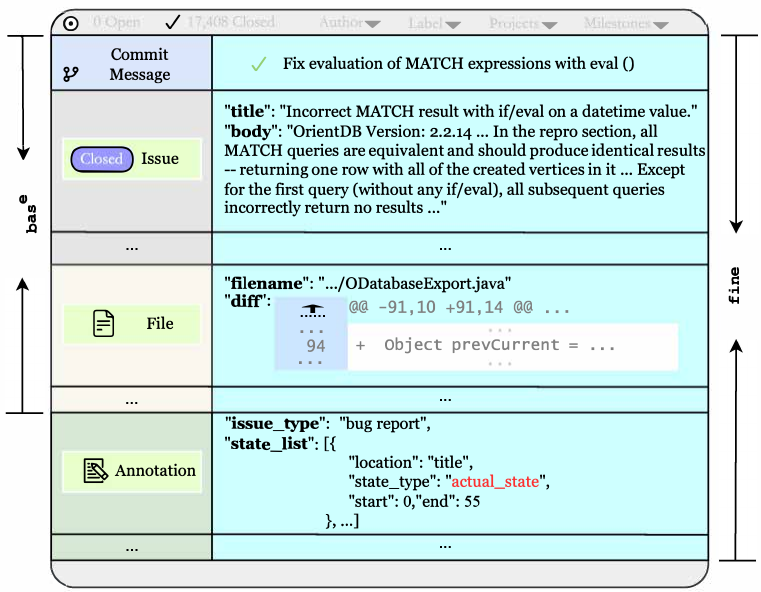}
    \caption{An example of the dataset}
    \label{data}
\end{figure}

\subsection{Example}
An example of the dataset is shown in Figure~\ref{data}. For each commit data in the base part of the dataset, we provide its commit message, a list of issues related to the commit, and a list of all files that have been modified in the commit. In the issue list, the title and body of each issue are given as strings. In the file list, the filename and code change for each file are also provided. For the fine part, in addition to containing all the information in the base part, it also includes annotation for each issue in the issue list. Specifically, we provide the type and the state information for each issue. The latter is represented by four key-value pairs: ``location'' indicates whether the information resides in the title or body; ``state\_type'' signifies whether the information is an actual state or an expected state; ``start'' and ``end'' respectively denote the offset of the starting and ending characters within the string.
\section{THE \tool PARADIGM}
\label{Sec.4}

\begin{figure*}[!ht] 
	\centering
	\includegraphics[width=1\linewidth]{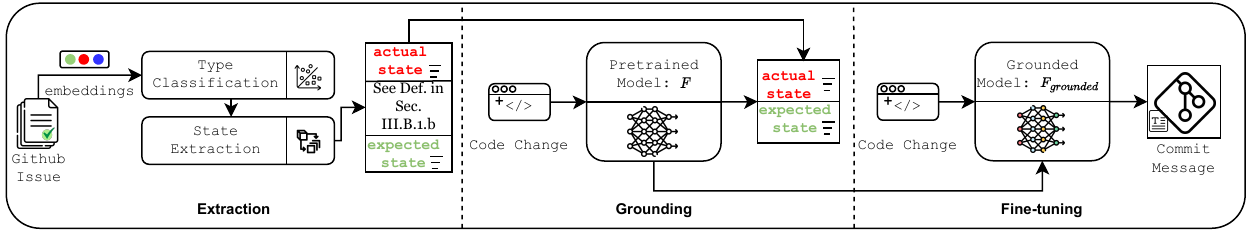} 
	\caption{Overview of \tool.}
	\label{egf}
\end{figure*}

In this section, we will introduce the proposed \tool (\underline{Ex}traction, \underline{Gro}unding, \underline{Fi}ne-tuning) paradigm, which exploits the correlation between commits and issues to generate more reasonable commit messages. The overview of \tool is illustrated in Figure~\ref{egf}. It is noteworthy that \tool is not a simple model that takes code changes as input and generates commit messages as output. Instead, it is a pipeline that integrates the fine-grained information carried by issues into a pre-trained CMG model to generate improved commit messages. The entire paradigm consists of three stages: extraction stage, grounding stage, and fine-tuning stage. During the first stage, we extract state information (defined in Sec.~\ref{state_info}) from existing issues. Then, in the grounding stage, we utilize the extracted information as target output and employ a sequence-to-sequence task to train a pre-trained CMG model $F$ to obtain a refined model $F_{grounded}$. Finally, in the fine-tuning stage, $F_{grounded}$ is trained to generate commit messages using code change. $F$ can be any existing learning-based or hybrid CMG model which uses sequentialized code change as input. After being trained with \tool, $F$ is transformed into $F_{grounded}$, which incorporates commit-issue correlation and possesses the ability to generate more rational commit messages. We then describe each stage of \tool in detail.

\subsection{Extraction Stage}
The purpose of this stage is to extract state information (defined in~\ref{state_info}) from a large corpus of issues. Through practical experimentation, we find that including the type of issue as a feature in the input enhances the extraction performance. Consequently, we design a pipeline structure within this stage: for each issue to be processed, we first employ a classifier to categorize its type, then input both the classification result and the issue text into a sequence labeling model~\cite{seqlab} to extract the state information. The internal structure of this stage is illustrated in Figure~\ref{extraction}.

\subsubsection{Issue Type Classification}
\label{itc}
For a given issue token list $T = \{t_{1}, t_{2}, ..., t_{n}\}$ ($n$ is the length of the sequence), we employ a pre-trained language model $PLM_{cls}$ to acquire the distributed representation of each token $H = \{h_{1}, h_{2}, ..., h_{n}\}$. Subsequently, the representation vector of the first token $h_{1}$ is fed into the classification head $MLP_{cls}$~\cite{mlp}, a linear layer with a \textit{softmax} activation function, to obtain a probability distribution across all categories. The category with the highest probability is identified as the type of input issue. Upon acquiring the type of the issue, we convey this information to the downstream model by concatenating a special token $t_{type}$ at the beginning of the input sequence. Specifically, the three types (bug report, feature request, and enhancement) are respectively represented by ``[BR]", ``[FR]", and ``[EN]". 

\subsubsection{State Information Extraction}
We model state information extraction as a sequence labeling (also known as sequence tagging) task~\cite{seqlab}. Sequence labeling is commonly employed to extract meaningful substrings from a lengthy character string, such as Named Entity Recognition (NER) task in natural language processing~\cite{ner_seqlab}. In the context of state information extraction, assuming that the AS (actual state) and ES (expected state) information is text within the issue, employing a sequence labeling approach for extraction is both reasonable and straightforward to implement. Under this modeling strategy, the model classifies each input token by assigning a tag. Different tags represent the positions of their corresponding tokens relative to the content to be extracted. We utilize the \textbf{BIO} labeling scheme~\cite{scheme}, which works as follows:
\begin{itemize}
    \item \textbf{B} (begin): This tag is assigned to the first token of the content to be extracted, and it is followed by the content type, e.g., B-AS, for the \underline{b}eginning of \underline{a}ctual \underline{s}tate information.
    \item \textbf{I} (inside): This tag is assigned to the tokens within the target content, also followed by the content type, e.g., I-ES for tokens \underline{i}nside \underline{e}xpected \underline{s}tate information.
    \item \textbf{O} (outside): This tag is assigned to tokens that are not part of any targeted content.
\end{itemize}
\begin{figure}[!b]
    \centering
    \includegraphics[width=1\linewidth]{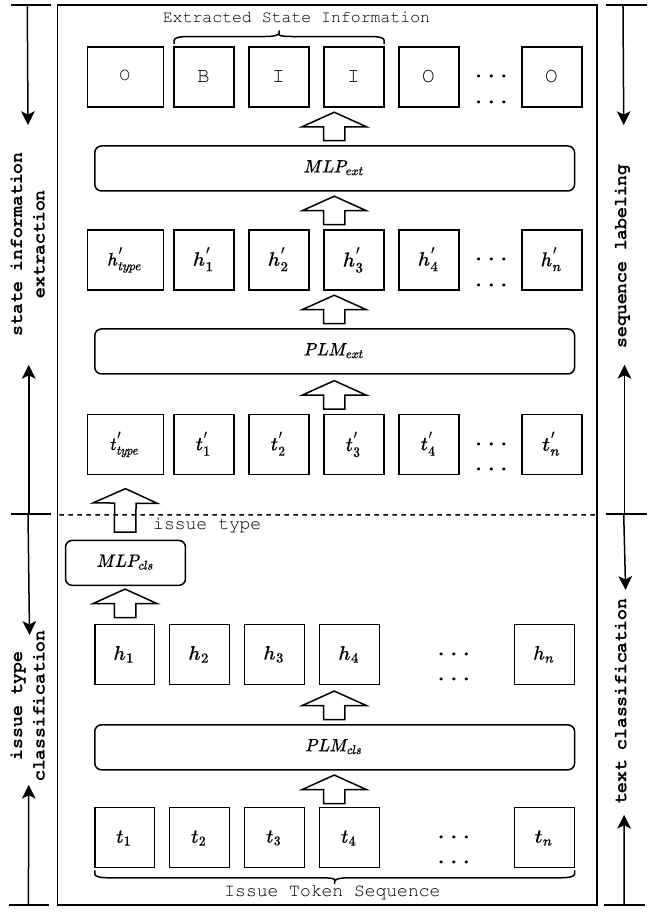}
    \caption{Extraction stage.}
    \label{extraction}
\end{figure}
Similar to the previous step, the state information extraction model comprises a pre-trained language model $PLM_{ext}$ and a linear classification head $MLP_{ext}$. The token sequence integrated with type information $T^{'} = \{t^{'}_{type}, t^{'}_{1}, t^{'}_{2}, ..., t^{'}_{n}\}$ is fed into the model to obtain hidden layer vectors in the semantic space $H^{'} = \{h^{'}_{type}, h^{'}_{1}, h^{'}_{2}, ..., h^{'}_{n}\}$. Subsequently, $MLP_{ext}$ performs classification on the hidden vectors at each position, yielding an output sequence composed of tags. Finally, the target information contained within the text can be decoded based on the schema. For each input sequence, the sum of the \textit{cross-entropy} for every token is considered as the final loss.

\subsection{Grounding Stage}
In the context of deep learning, grounding~\cite{grounding} refers to connecting abstract representations learned by a neural network to real-world knowledge. In this paper, we draw inspiration from this terminology and refer to the process of enhancing the model's capacity to represent code change as grounding. During the grounding stage, we aim to embed the code change into a representation vector that more accurately reflects its semantics. According to the definition of state information in Sec.~\ref{Sec.3.2.1}, issues can offer descriptions of the state that the code should exhibit before and after the change. In other words, the state information can be regarded as a natural language summary of the code change. Therefore, it can serve as a supervisory objective for code change representation learning.

The grounding stage can be initialized by any pre-trained commit message generation model $F$, which takes code diff as input. Given a sequentialized code diff token sequence, the objective of this stage is to obtain a grounded model $F_{grounded}$ that can align input code change with words possessing the same semantic information. By concatenating state information of related issues as the target language, we can model the training process of $F$ as a sequence-to-sequence task~\cite{seq2seq}. Since the model outputs a probability distribution over the vocabulary on all translation steps, the summed \textit{cross-entropy} loss is also used as the objective function. After grounding, we acquire a model $F_{grounded}$, which possess the knowledge learned from the correlation between commits and issues. 

\subsection{Fine-tuning Stage}
The fine-tuning stage is relatively simple: Given an input sequence of code change, $F_{grounded}$ is fine-tuned to generate a commit message in this stage. Note that the structure of $F_{grounded}$ and $F$ are the same, but the parameters are different. 
\section{EXPERIMENTAL SETUP}
\label{Sec.5}
\subsection{Research Question}
\begin{itemize}
    \item \textbf{RQ1: Overall effectiveness.} To what extent can \tool enhance the performance of existing CMG models?
    \item \textbf{RQ2: Extraction performance.} How effective does the first stage of \tool extract state information from issues?
    \item \textbf{RQ3: Contribution of grounding stage.} Do models really learn useful correlation knowledge through the grounding stage?
    \item \textbf{RQ4: Human evaluation.} How do \tool-enhanced models perform from the perspective of human evaluation?
\end{itemize}

\vspace{-5pt}

\subsection{Data}
Since there was no commit-issue parallel corpus before, we use the dataset described in Section~\ref{Sec.3} for all experiments. The annotated issues in the fine part are used to evaluate models in the extraction stage. The base part is used to evaluate the grounding stage and the generation stage.

\subsection{Enhanced Models}
We use \tool to enhance six state-of-the-art models as follows.
\begin{itemize}
    \item \textbf{Encoder-only models} only provide contextualized code diff representation by a pre-trained encoder. We consider three encoder-only pre-trained models CodeBERT~\cite{codebert}, CodeBERTa~\cite{codeberta}, and UniXcoder~\cite{unixcoder}.
    \item \textbf{Encoder-decoder models} possess both pre-trained encoder for representation and decoder for generation. We consider two encoder-decoder models CodeTrans~\cite{codetrans} and CodeT5~\cite{codet5}. We experimented with both the SMALL and BASE versions of these two models.
    \item \textbf{Hybrid models} use both retrieval-based and learning-based techniques. We select RACE~\cite{race}, which is a retrieval-augmented neural commit message generation model.
\end{itemize}
Note that models using the ASTs of the code (such as FIRA~\cite{fira} and ATOM~\cite{atom}) do not participate in our evaluation. This is because the code diff fragments provided in the dataset we used are unparseable.

\subsection{Implementation}
\paragraph{Models}
We use the official parameters downloaded from Hugging Face~\cite{hf} to initialize our models. For encoder-only and encoder-decoder models, we directly use the downloaded pre-trained checkpoint. But encoder-only models do not have decoders for generation. So we randomly initialize a Transformer~\cite{vaswani2017attention} decoder for encoder-only models. The hyperparameters of these random decoders are set according to the configuration of the corresponding encoder model. For RACE, we initialized it with the weight of CodeT5-base as reported in its original paper. In addition, for the pre-trained language model in the extraction stage, we use CodeBERT.
\paragraph{Tokenizers}
As mentioned in Sec.~\ref{itc}, we add three special tokens ``[BR]", ``[FR]", and ``[EN]" for the tokenizer of extraction stage to indicate the type of issue type.
\paragraph{Environments}
The experiments are performed on an NVIDIA DGX Station with Intel Xeon CPU E5-2698 v4 @ 2.2 GHz, running Ubuntu 20.04.4 LTS. The models are trained on one 32G GPU of NVIDIA TESLA V100.

\subsection{Automatic Metrics}
\subsubsection{Commit Message Generation}
Numerous automated metrics have been employed for the evaluation of the CMG task. However, each evaluation metric possesses inherent limitations. Consequently, in order to assess \tool from diverse perspectives, we have selected the following four distinct automated metrics for our analysis.
\begin{itemize}
    \item \textbf{BLEU}~\cite{papineni2002bleu} measures the precision of n-grams (sequences of n words) in the generated text compared to the reference texts. It calculates the modified n-gram precision and applies a brevity penalty to avoid favoring shorter sentences.
    \item \textbf{ROUGE-L}~\cite{lin2004rouge} is a metric that evaluates the quality of summaries by measuring the longest common subsequence (LCS)~\cite{paterson1994longest} between the generated text and the reference texts. It is less sensitive to the exact word order than BLEU but provides a more recall-oriented evaluation.
    \item \textbf{METEOR}~\cite{banerjee2005meteor} calculates the harmonic mean of unigram precision and recall between the generated text and the reference text, considering exact matches, synonyms, and stemmed forms of words.
    \item \textbf{CIDEr}~\cite{vedantam2015cider} computes the Term Frequency-Inverse Document Frequency (TF-IDF) weighting for each n-gram to emphasize the importance of informative and rare n-grams. Therefore, it is more effective in capturing the relevance of specific details.
\end{itemize}

\begin{table*}[!t]
\centering
\captionsetup{justification=centering}
\caption{Generation performance for \tool-enhanced commit message generation models.}
\resizebox{1\linewidth}{!}{
\begin{tabular}{@{}lcccccccccccc@{}}
\toprule
\rowcolor[HTML]{FFFFFF} 
\cellcolor[HTML]{FFFFFF} &
  \multicolumn{3}{c}{\cellcolor[HTML]{FFFFFF}BLEU} &
  \multicolumn{3}{c}{\cellcolor[HTML]{FFFFFF}ROUGE-L} &
  \multicolumn{3}{c}{\cellcolor[HTML]{FFFFFF}METEOR} &
  \multicolumn{3}{c}{\cellcolor[HTML]{FFFFFF}CIDEr} \\ \cmidrule(l){2-13} 
\rowcolor[HTML]{FFFFFF} 
\multirow{-2}{*}{\cellcolor[HTML]{FFFFFF}\textbf{Model}} &
  \cellcolor[HTML]{FFFFFF}Ori. &
  \cellcolor[HTML]{FFFFFF}Enh. &
  \cellcolor[HTML]{FFFFFF}Imp. &
  Ori. &
  Enh. &
  Imp. &
  Ori. &
  Enh. &
  Imp. &
  Ori. &
  Enh. &
  Imp. \\ \midrule
\rowcolor[HTML]{FFFFFF} 
CodeBERT~\cite{codebert} &
  8.81 &
  9.09 &
  \cellcolor{black!25}3\% &
  15.85 &
  16.05 &
  \cellcolor{black!25}1\% &
  4.75 &
  5.0 &
  \cellcolor{black!25}5\% &
  0.09 &
  0.11 &
  \cellcolor{black!25}22\% \\
\rowcolor[HTML]{FFFFFF} 
CodeBERTa~\cite{codeberta} &
  8.43 &
  8.50 &
  \cellcolor{black!25}1\% &
  14.86 &
  15.66 &
  \cellcolor{black!25}5\% &
  4.92 &
  5.12 &
  \cellcolor{black!25}4\% &
  0.09 &
  0.11 &
  \cellcolor{black!25}22\% \\
\rowcolor[HTML]{FFFFFF} 
unixcoder~\cite{unixcoder} &
  9.46 &
  9.54 &
  \cellcolor{black!25}1\% &
  17.04 &
  17.59 &
  \cellcolor{black!25}3\% &
  5.57 &
  6.18 &
  \cellcolor{black!25}11\% &
  0.14 &
  0.16 &
  \cellcolor{black!25}14\% \\ \midrule
\rowcolor[HTML]{FFFFFF} 
\textbf{encoder-only average} &
  8.9 &
  9.04 &
  \cellcolor{black!25}2\% &
  15.92 &
  16.43 &
  \cellcolor{black!25}3\% &
  5.08 &
  5.43 &
  \cellcolor{black!25}7\% &
  0.11 &
  0.13 &
  \cellcolor{black!25}18\% \\ \midrule
\rowcolor[HTML]{FFFFFF} 
CodeTrans-small~\cite{codetrans} &
  11.08 &
  12.66 &
  \cellcolor{black!25}14\% &
  18.62 &
  20.84 &
  \cellcolor{black!25}12\% &
  7.98 &
  9.07 &
  \cellcolor{black!25}14\% &
  0.3 &
  0.46 &
  \cellcolor{black!25}53\% \\
\rowcolor[HTML]{FFFFFF} 
CodeTrans-base~\cite{codetrans} &
  19.27 &
  20.37 &
  \cellcolor{black!25}6\% &
  27.34 &
  28.53 &
  \cellcolor{black!25}4\% &
  16.37 &
  18.31 &
  \cellcolor{black!25}12\% &
  1.0 &
  1.08 &
  \cellcolor{black!25}8\% \\
\rowcolor[HTML]{FFFFFF} 
CodeT5-small~\cite{codet5} &
  11.40 &
  12.59 &
 \cellcolor{black!25}10\% &
  20.70 &
  22.68 &
  \cellcolor{black!25}10\% &
  8.86 &
  10.02 &
  \cellcolor{black!25} 13\% &
  0.36 &
  0.49 &
  \cellcolor{black!25} 37\% \\
\rowcolor[HTML]{FFFFFF} 
CodeT5-base~\cite{codet5} &
  19.96 &
  20.94 &
  \cellcolor{black!25}5\% &
  30.07 &
  31.46 &
  \cellcolor{black!25}5\% &
  16.99 &
  18.39 &
  \cellcolor{black!25}8\% &
  1.04 &
  1.11 &
  \cellcolor{black!25}7\% \\
\rowcolor[HTML]{FFFFFF} 
RACE~\cite{race} &
  21.06 &
  22.47 &
  \cellcolor{black!25} 7\% &
  30.17 &
  36.09 &
  \cellcolor{black!25} 20\% &
  17.23 &
  18.64 &
  \cellcolor{black!25} 8\% &
  1.14 &
  1.51 &
  \cellcolor{black!25} 32\% \\ \midrule
\rowcolor[HTML]{FFFFFF} 
\textbf{\begin{tabular}[c]{@{}l@{}}encoder-decoder + hybrid\\ average\end{tabular}} &
  16.55 &
  17.81 &
  \cellcolor{black!25} 8\% &
  25.38 &
  27.92 &
  \cellcolor{black!25} 10\% &
  13.486 &
  14.886 &
  \cellcolor{black!25} 10\% &
  0.768 &
  0.93 &
  \cellcolor{black!25}21\% \\ \bottomrule
\end{tabular}}
\begin{tablenotes}
\footnotesize
    \item ``\textbf{Ori.}" means Original; ``\textbf{Enh.} " is Enhanced; ``\textbf{Imp.}" indicates Improvement.
\end{tablenotes}
\label{rq1_table}

\end{table*}

\subsubsection{State Information Extraction}
Drawing upon the information extraction domain~\cite{ner_re}, we opted to use precision, recall, and F1-score (the harmonic mean of precision and recall ) to evaluate the performance of the extraction stage. During the actual training of the state information extractor, we utilized the fine part of the dataset, which was randomly divided into an 8:1:1 ratio for the training, validation, and testing sets. A micro-F1 score calculated over the whole test set is also presented. It is essential to note that the information we aim to extract is typically longer strings, so requiring the model's output to be identical to the golden answer might be excessively stringent. Therefore, when calculating the metrics, we employ a fuzzification strategy to assess whether the extracted content by the model meets the requirements. Specifically, we first compute the number of matching character between the model output string and the golden string. Then we calculate a similarity score by dividing the number of matching characters by the average length of the strings. If the score exceeds a predetermined threshold $\tau$ (we set the value as 0.8), we consider the two strings to match, indicating that the model has extracted the correct answer.

\section{RESULTS AND ANALYSIS}
\label{Sec.6}
In this section, we present the overall effectiveness of \tool-enhanced models for CMG (RQ1) in Sec.~\ref{rq1}, the performance of the extraction stage (RQ2) in Sec.~\ref{rq2}, the contribution of grounding stage (RQ3) in Sec.~\ref{rq3} and human evaluation (RQ4) in Sec.~\ref{rq4}.

\subsection{RQ1: Overall Effectiveness}
\label{rq1}

The overall effectiveness of \tool{} is presented in Table~\ref{rq1_table}.
It can be observed that the generation performance of each model is improved after being trained using \tool. Among them, the overall generation performance of encoder-only models is relatively inferior, which can be attributed to their decoders being initialized randomly. In contrast, the generation performance of pre-trained encoder-decoder models is considerably superior. Regardless of whether the models are initialized entirely with pre-trained parameters, those trained using \tool consistently outperform models that solely rely on fine-tuning. This demonstrates the effectiveness of \tool in the CMG task. Specifically, for encoder-only models, \tool can increase the average scores of BLEU, ROUGE, METEOR, and CIDER by 2\%, 3\%, 7\%, and 18\%, respectively. For encoder-decoder models, these four metrics can be improved by 8\%, 10\%, 10\%, and 21\%, respectively. The most notable improvement is observed in CIDEr, indicating that models trained with \tool are more likely to generate outputs that capture specific details or content deemed significant by humans. This is crucial in the scenario of CMG task~\cite{what_why}, as different commit messages often contain vocabulary which is unique to specific code repositories, such as particular class names and variable names. In addition, being limited by input length, original models can not fully cover all information in the code change. But with the aid of \tool, models can obtain knowledge from issues, significantly reducing the loss of information when representing code change and consequently producing more rational commit messages.

\find{{\bf \ding{45} Answer to RQ-1: }$\blacktriangleright$ 
\tool{} improves generation performance in the CMG task for encoder-only, encoder-decoder and hybrid models, with encoder-decoder models showing superior results. Notable improvements are observed in CIDEr, indicating better capture of specific human-relevant details. \tool also aids in reducing information loss, enabling more rational commit messages.
$\blacktriangleleft$ }

\begin{table}[!b]
\centering
\captionsetup{justification=centering}
\caption{State information extraction performance}
\vspace{0.05cm}   
\begin{tabular}{ll|lr}
\toprule
\multicolumn{2}{l|}{\textbf{Approach}} & codebert & + issue type   \\ 
\midrule
\multicolumn{1}{l|}{\multirow{3}{*}{\textbf{\begin{tabular}[c]{@{}l@{}}Actual\\ State\end{tabular}}}}   & \textbf{P} & 56.90 & \textbf{58.93} \\
\multicolumn{1}{l|}{}   & \textbf{R}   & 63.46    & \textbf{64.71} \\
\multicolumn{1}{l|}{}   & \textbf{F1}  & 60.00    & \textbf{61.68} \\ \hline
\multicolumn{1}{l|}{\multirow{3}{*}{\textbf{\begin{tabular}[c]{@{}l@{}}Expected\\ State\end{tabular}}}} & \textbf{P} & 80.77 & \textbf{84.91} \\
\multicolumn{1}{l|}{}   & \textbf{R}   & 72.41    & \textbf{77.59} \\
\multicolumn{1}{l|}{}   & \textbf{F1}  & 76.36    & \textbf{81.08} \\ \hline
\multicolumn{2}{l|}{\textbf{Micro-F1}} & 68.18    & \textbf{71.56} \\
\bottomrule
\end{tabular}

\label{rq2_table}
\end{table}

\subsection{RQ2: Extraction Performance}
In the extraction stage of \tool, a learning-based model is employed to extract state information, necessitating an evaluation of its performance. During the training phase of the state information extractor, we utilized the fine part of the dataset, which was randomly divided into an 8:1:1 ratio for the training, validation, and testing sets. In Table~\ref{rq2_table}, we present the precision (P), recall (R), and F1-score for the two categories of state information, as well as the micro-F1 for the entire test set. The middle column and the rightmost column of the table represent the extraction performance without and with the utilization of category information, respectively. It can be observed that the model's extraction performance exhibits a significant improvement when incorporating category information. Furthermore, the model demonstrates a better extraction capability for expected state information (F1-score 81.08\%) compared to actual state information (F1-score 61.68\%). This is primarily due to the fact that expected state is often proposed by more experienced developers or users~\cite{bettenburg2008makes}, resulting in higher data consistency than the actual state, which consequently makes it easier for the model to extract. However, the overall micro-F1 score reaches 71.56\%. Based on the experience in the information extraction domain~\cite{ner_re}, this performance is sufficient to meet the demands of state information extraction for data in our scenario.

\label{rq2}
\begin{figure}[!b] 
	\centering
	\includegraphics[width=1\linewidth]{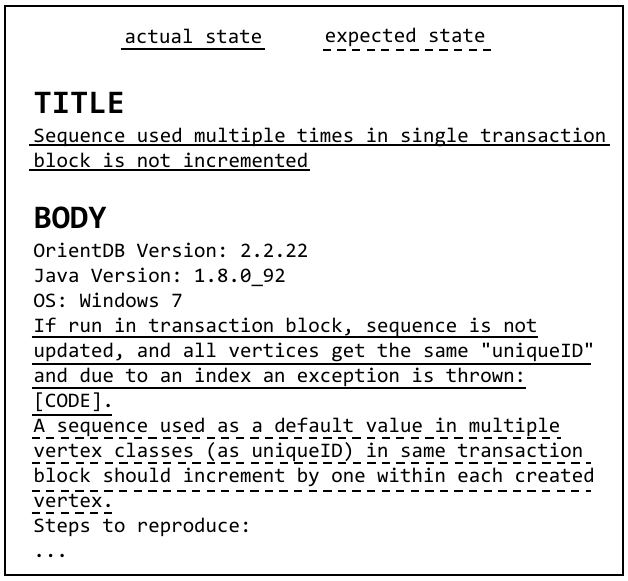} 
	\caption{Example of extracted state information.}
	\label{rq2_fig}
\end{figure}

Figure~\ref{rq2_fig} provides an example of the extracted state information from a real issue. In the figure, we show both the title and the body of this issue. Actual state information is indicated by solid underlines, while expected state information is indicated by dashed underlines. The former describes the actual behavior which is unexpected, and the latter suggests the expected behavior.

\find{{\bf \ding{45} Answer to RQ-2: }$\blacktriangleright$ 
In the extraction stage of \tool{}, incorporating category information significantly improves state information extraction performance, with better results for expected state (F1-score 81.08\%) than actual state (F1-score 61.68\%). The overall micro-F1 score reaches 71.56\%, sufficient to meet state information extraction demands for the scenario.
$\blacktriangleleft$ }

\subsection{RQ3: Contribution of Grounding Stage}
\label{rq3}
To validate whether the grounding stage of \tool can integrate knowledge related to commits from issues into models, we conduct a comparative experiment. Firstly, we extract the encoders from the original CodeT5-base and the \tool-enhanced CodeT5-base. Then, we utilize these two encoders to individually embed the code changes and commit messages of each data example in the test set of our dataset, resulting in two pairs of vectors. We also standardize these vector values to eliminate the influence of dimension. Then we compute the \textit{Euclidean distance} between the vectors in each pair, representing the distance in the semantic space. A smaller distance implies a higher semantic similarity between the obtained code change representations and commit message representations. Figure~\ref{distance} presents the distribution for pairs embedded by encoders before and after grounding. The results show that the representations provided by the grounded encoder are more similar. The Mann-Whitny-Wilcoxon test~\cite{mcknight2010mann} (p-value: 1.9e-55) further validates the significance of the difference before and after grounding. This indicates that \tool enables the model to more accurately align code changes within the semantic space, thereby generating more reasonable commit messages.

\find{{\bf \ding{45} Answer to RQ-3: }$\blacktriangleright$ 
The grounding stage of \tool effectively integrates commit-related knowledge from issues. A comparative experiment shows significantly smaller Euclidean distances between code change and commit message representations in \tool-enhanced models, indicating better semantic alignment and more reasonable commit message generation.
$\blacktriangleleft$ }

\subsection{RQ4: Human Evaluation}
\label{rq4}
While automatic metrics can provide valuable insights, the lack of semantic understanding still makes them insufficient for a comprehensive evaluation. Therefore, following Shi et al.~\cite{race}, we conduct a human evaluation to further study the quality of commit messages generated by \tool-enhanced models. We randomly select 50 data samples from the test set of the base part and design a questionnaire for evaluation. For each sampled data, the questionnaire includes the code change, the ground truth commit message, the commit message generated by the original model and the commit message generated by the \tool-enhanced model. We invited four experienced developers to conduct human evaluation, including two software engineers with over two years of work experience and two graduate students with long-term internship experience. Each evaluator needs to assess the quality of the generated commit messages from the following four aspects.: 
\begin{itemize}
    \item \textbf{Rationality}: Whether it provides a logical explanation for the changes, addressing the ``why" behind the commit.
    \item \textbf{Comprehensiveness}: Whether it covers all important details and provides a complete picture of the modifications made, addressing the ``what" behind the commit.
    \item \textbf{Conciseness}: Whether it conveys information succinctly, ensuring readability and quick comprehension.
    \item \textbf{Expressiveness}: Whether its content is grammatically correct and fluent.
\end{itemize}

For each aspect, the developer must indicate whether the original or the enhanced model's generated result is better. If the enhanced model is better, they assign ``Win"; otherwise, assign ``Lose". If they think the difference between the two is not obvious, they should assign a ``Tie" label. To mitigate bias, each developer fills in the questionnaire independently and the agreement among them is measured by Fleiss' kappa~\cite{kappa}.

\begin{figure}[!t] 
	\centering
	\includegraphics[width=1\linewidth]{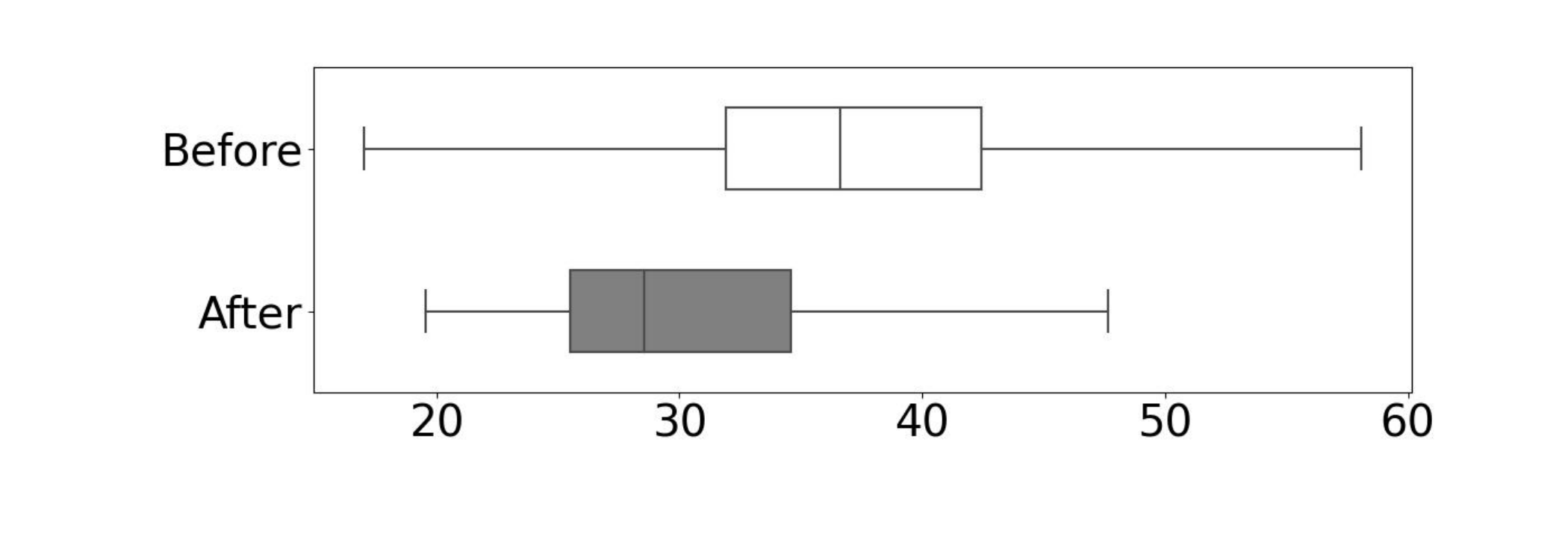} 
	\caption{Euclidean distance between embedding of code change and commit message before and after grounding.}
	\label{distance}
\end{figure}

\begin{table}[!t]
\centering
\caption{Result of human evaluation.}

\label{rq4_table}
\resizebox{1\linewidth}{!}{\begin{tabular}{l|cccc}
\toprule
\textbf{Indicator} & Win (N-\%)                & Lose (N-\%)     & Tie (N-\%)      & Kappa \\ \midrule
\textbf{Rat.}       & \cellcolor{black!25}35 (70\%) & 10 (20\%) & 5 (10\%)  & 0.70  \\
\textbf{Comp.}  & \cellcolor{black!25}25 (50\%) & 14 (28\%) & 11 (22\%) & 0.67  \\
\textbf{Conc.}      & \cellcolor{black!25}21 (42\%) & 20 (40\%) & 9 (18\%)  & 0.65  \\
\textbf{Expr.}    & \cellcolor{black!25}17 (34\%)          & 17 (34\%) & 16 (32\%) & 0.66 \\
\bottomrule
\end{tabular}}
\begin{tablenotes}
\footnotesize
    \item \textbf{N} is short for number; \% indicates percentage.
    \item \textbf{Rat.} means Rationality. \textbf{Comp.} is Comprehensiveness.
    \item \textbf{Conc.} is Conciseness. \textbf{Expr.} represents Expressiveness.
    
\end{tablenotes}
\end{table}

Table~\ref{rq4_table} reports the results of human evaluation. The values of kappa are all above 0.6, indicating substantial agreement among the four developers. According to the results, we notice that \tool significantly improves upon rationality while achieving comparable performance on comprehensiveness, conciseness and expressiveness, which substantiates that our proposed paradigm indeed enables models to learn rational information from the correlation between commits and issues.

After human evaluation, we also select three examples that intuitively demonstrate the enhancement of the model by the tool, as shown in Table~\ref{cases}. Each example not only provides the generated results of the original model and the \tool-enhanced model, but also provides partial text from the issue related to that commit. It can be seen that, the commit message generated by the original model only provides a rough description of the modifications. In comparison, the \tool-enhanced model's output includes specific descriptions of the modified objects (e.g.,``handler in Mono.subscribe()" in Example 2 and``when starting process instance" in Example 3) as well as the reasons for the modifications (e.g.,``to reclaim the memory" in Example 1 and ``Fix bug" in Example 3). These examples further demonstrate that \tool, by introducing correlation knowledge between commits and issues, enables models to have the ability to generate more rational results.

\vspace{15pt}

\begin{table*}[!b]
\centering
\captionsetup{justification=centering}
\caption{Examples that demonstrate \tool{}'s effect.}
\resizebox{0.9\linewidth}{!}
{
\begin{tabular}{l|l|l}
\hline
                             & \textbf{Before}                  & Add support for custom field.                                                                                \\
                             & \textbf{After}                   & Fix TimingKey \textbf{to reclaim the memory}.                                                                       \\
\multirow{-3}{*}{\textbf{1}} & \textbf{Issue}                   & ... {\color[HTML]{FE0000} There is no mechanism to reclaim the memory after the filters are unreachable} ... \\ \hline
                             & \textbf{Before}                  & Reduce visualisation to fix issue.                                                                           \\
                             & \textbf{After}                   & Fix exception with error \textbf{handler in Mono.subscribe()}.                                                        \\
\multirow{-3}{*}{\textbf{2}} & \textbf{Issue}                   & ... {\color[HTML]{32CB00} Error handler in Mono.subscribe() should handle exceptions thrown by subscriber} ... \\ \hline
                             & \textbf{Before}                  & Update process instance.                                                                                     \\
                             & \textbf{After}                   & \textbf{Fix bug}  \textbf{when starting process instance}.                                                                      \\
\multirow{-3}{*}{\textbf{3}} &                                  & ... {\color[HTML]{FE0000} No tenant is set when starting process instance }...                                                      \\
                             & \multirow{-2}{*}{\textbf{Issue}} & ... {\color[HTML]{32CB00} Expect a tenant user can start a process instance in task application } ...                                \\ \hline
\end{tabular}
}

\begin{tablenotes}
\footnotesize
    \item ``\textbf{Before}" and ``\textbf{After}" represent the commit messages generated by the model before and after using \tool{}, respectively.
    \item ``\textbf{Issue}" provides part of the text in the related issue; among them, the {\color[HTML]{FE0000} red text} indicates the actual state, and the {\color[HTML]{32CB00} green text} indicates the expected state.
\end{tablenotes}

\label{cases}
\end{table*}

\find{{\bf \ding{45} Answer to RQ-4: }$\blacktriangleright$ 
Human evaluation of commit messages generated by \tool{}-enhanced models shows significant improvement in rationality while maintaining comparable performance in comprehensiveness, conciseness, and expressiveness. This confirms that \tool{} effectively enables models to learn rational information from the correlation between commits and issues.
$\blacktriangleleft$ }

\section{Discussion}
\label{Sec.7}

We discuss the threats to validity and enumerate a few limitations of our study.

\subsection{Threats to Validity}

The internal threat to validity lies in the implementation of approaches and the setting of hyperparameters. To reduce this threat, we directly reuse the open-source implementation of RACE~\cite{race} and public checkpoints of all pre-trained models on Hugging Face~\cite{hf}. To ensure the optimal hyperparameters for each model, we conducted numerous iterative experiments and tested models with various decoding parameter configurations to generate the best possible results.

The external threat to validity lies in the source of data that we used to construct the dataset. To mitigate this threat, we employed the official GitHub REST API~\cite{githubrestapi} to collect data from the top 1,000 repositories ranked by the number of stars. Furthermore, referencing previous research~\cite{nngen, what_why, mcmd}, we conducted extensive work in cleaning, filtering, and processing the data to ensure the highest possible quality.

The threat to construct validity arises from the evaluation metrics employed. To mitigate this threat, we utilize four metrics that have been extensively employed in previous studies on commit message generation~\cite{nngen, race}. Moreover, we conduct a human evaluation to assess the effectiveness from the developers' perspective. By rigorously adhering to the methodology of prior research and engaging experienced developers, we aim to minimize potential threats in human evaluation.

\subsection{Limitations}

This section discusses the limitations of our work. Firstly, the amount of annotated data utilized for training the extraction module is relatively limited. Although the current volume enables the module to possess a satisfactory extraction capability, additional data would undoubtedly improve the extraction performance. Secondly, due to the pipeline structure within \tool, error propagation exists between different stages. The quality of issue data and the performance of the upstream stage both influence the final generation outcome. Lastly, although the results generated by \tool-enhanced models contain more rational information, these commit messages are sometimes not concise enough.
\section{RELATED WORK}
\label{Sec.8}
Existing methods for commit message generation can be categorized as template-based, learning-based, information retrieval-based and hybrid models.

Early methods were typically template-based~\cite{buse2010automatically, cortes14, shen2016automatic}. These techniques used parsers to analyze the modification of source code and generated commit messages with pre-defined rules. For instance, Buse et al.~\cite{buse2010automatically} symbolically executed code changes to acquire path predicates, then generated commit message using a set of pre-defined rules based on the path. Cortés-Coy et al.~\cite{cortes14} proposed a template based on stereotypes~\cite{dragan2011using, dragan2006reverse}, then filled the template with extracted information to generate commit message. 

Learning-based methods~\cite{commitgen, nmt, codisum, ptr, fira} drew inspiration from the idea of neural machine translation and modeled the generation of commit message as a sequence-to-sequence task. Jiang et al.~\cite{commitgen} early attempted to respectively treat code change and commit message as input and output. Loyola et al.~\cite{nmt} proposed an encoder-decoder model with the attention mechanism proposed by Luong et al.~\cite{luong-etal-2015-effective}. Dong et al.~\cite{fira} represented code change via fine-grained graphs and generated commit message with graph neural network~\cite{wu2020comprehensive}.

Retrieval-based approaches~\cite{nngen, cc2vec} sorted existing commit messages according to similarity as output. Liu et al.~\cite{nngen} represented code change as bag-of-words vectors and then used the nearest neighbor algorithm to sort. Hoang et al.~\cite{cc2vec} learned a distributed representation for code change guided by commit message. The learned representations are used to adapt the bag-of-words vectors of the former model.

Hybrid architectures~\cite{atom} combined the former two types of approaches. Liu et al. made use of the Abstract Syntax Trees of the code change in the learning module and sorted them according to cosine similarity in the retrieval module. The final result is output by a ranking module that prioritize the commit message obtained from the former two modules.

Influenced by powerful pre-trained language models in natural language processing~\cite{bert},~\cite{li2021pretrained}, many pre-trained models that provide distributed representations of code have emerged in recent years~\cite{codebert},~\cite{codetrans},~\cite{codet5}. At present, the main paradigm to generate commit message is to combine pre-trained code representation models with task-specific generation modules~\cite{race}.

\section{CONCLUSION AND FUTURE WORK}
\label{Sec.9}

In this work, we delve into the correlation between commits and issues, addressing a previously unexplored aspect of automatic commit message generation. We collect a large commit-issue parallel dataset, allowing for a deeper understanding of the rational information issues provide. A novel paradigm \tool is also proposed to extract and incorporate fine-grained knowledge from issues into the commit message generation process. The extensive evaluation demonstrates the effectiveness of \tool in significantly improving the rationality of generated commit messages while maintaining other performance indicators. 

This paper can serve as a foundation for future research in commit message generation, particularly in exploring methods of extracting and incorporating external information from issues and other modalities. For instance, researchers can explore the pre-training of large-scale language models based on such commit-issue correlation, investigate more efficient methods for incorporating external knowledge into the commit message generation process, or transfer principles of \tool to other tasks in the field of automated software engineering.

\section{Artifact Availability}
We make our code and dataset publicly available at:

    {\url{https://anonymous.4open.science/r/ExGroFi-FA21}}
    
\section{Acknowledgments}
This work is supported by the National Natural Science Foundation of China (Grant Nos. 62276017, U1636211, 61672081), the Fund of the State Key Laboratory of Software Development Environment (Grant No. SKLSDE-2021ZX-18) and the NATURAL project which has received funding from the European Research Council (ERC) under the European Union’s Horizon 2020 research and innovation programme (Grant No. 949014).

\bibliographystyle{IEEEtran}
\bibliography{ase2023}

\end{document}